\documentstyle[12pt]{article}
\advance\leftskip by -20pt \advance\rightskip by -25pt
\advance\vsize by 4.0 truecm
\advance\voffset by -1.0 truecm
\begin{document}
\baselineskip18pt
\begin{titlepage}
\date{}
\title{Induced gravity inflation in the SU(5) GUT}
\author{J.L.Cervantes-Cota\thanks{e-mail:
jorge@spock.physik.uni-konstanz.de} and H. Dehnen \thanks{e-mail:
dehnen@spock.physik.uni-konstanz.de} \\[10pt]
    University of Konstanz, p.o.box: 5560 M 678
    \,\ D-78434 Konstanz }
\maketitle
\vskip 1 cm
To appear in Phys. Rev.D, 15th. January 1995
\vspace{1cm}
\thispagestyle{empty}
\begin{abstract}
\baselineskip18pt

We investigate the cosmological consequences of a theory of induced
gravity in which the scalar field is
identified with the Higgs field of the  first
symmetry breaking  of a mi\-ni\-mal SU(5) GUT.  The
mass of the X-boson
determines a great value for the coupling constant of gravity-particle
physics.  Because of this fact, a "slow"
rollover dynamics for the Higgs
field is not possible in a "new" inflation scenario and,   moreover,  a
contraction era for the scale factor in the early universe exists, \-after\-
which inflation follows automatically; "chaotic" inflation is performed
without problems.   Inflation is successfully achieved due to the
relationship among the masses of particle physics at that scale: The Higgs-,
X-boson- and Planck-masses. As a result the particle physics
pa\-ra\-me\-ter $\lambda$ is not fine-tuned as usual in order to predict
 acceptable
values of re-heating temperature and density and  gravitational wave
perturbations.  Moreover, if the coherent Higgs oscillations didn't
decay they
could explain the missing mass problem of cosmology.
\end{abstract}
\end{titlepage}

\newpage
\section{INTRODUCTION}
It was more than a decade ago that the inflationary model \cite{Gu81}
was proposed in order to solve some problems in cosmology:
the horizon and flatness problems, the present isotropy of
the universe and the possible overabundance
of magnetic monopoles.  The
inflationary scenario was inspired to incorporate
theories
of particle physics into the early universe,  achieving an interesting joint
of these two different areas of physics.  Since that time
there were a lot of alternative models
(for a review see refs. \cite{KoTu90,Li90,Ol90,GoPi92})
formulated to overcome the
problems that inflation suffers, {\it i.e.}, a smooth ending of the
inflationary era
(graceful exit), enough e-folds of inflation,
sufficient reheat temperature
for baryogenesis to take place and the right contrast of density and
gravitational perturbations coming from the scalar field fluctuations, among
others, to achieve successfully inflation \cite{StTu84}.  Although some
inflationary scenarios can solve many of the above mentioned
 problems, some
other features of them (coupling constant's strengths, etc.) are
not well understood, just as a consequence of
the lack of a final gravity theory  coupled to the other
interactions of nature,
necessary to describe the very early universe.  Ne\-ver\-the\-less, gravity
theories including  quadratic terms  coming from high energy theories or
Brans-Dicke Theory (BDT) plus potentials, induced gravity
theories, or others with a particle content are today believed
to be the more realistic ones in order to describe the first stage  of our
universe and, according with the experimental constraints,
for low energies these theories should be not much different
from Einstein's General Relativity (GR) in a 4-spacetime manifold.

There have been considered many Lagrangians in cosmology inspired
from particle physics that should be applicable at the beginning
of the universe.  Some of the scenarios coming from these Lagrangians
are not very carefully in treating the
coupling gravity-particle physics, but
they are in adjusting the particle physics parameters to the
 cosmological
ones in order to solve the above known problems of cosmology,
 resulting in a
necessary but undesirable fine-tuning.  In particular, to obtain
the right contrast of density perturbations caused by preinflationary
field fluctuations, one has to set $\lambda$
(from an inflation potential, say, $ V=\lambda \phi^{4} $) to a very small
number  ($< 10^{-12} $) by hand, which is
from the particle physics point of
view  very unnatural  and has no justification.

The cosmological consequences of induced gravity models are well
known \cite{Sp84,AcZoTu85,Po85,LuMaPo86,Po86,AcTr89,FaUn90},  but the
 particle physics content is still unclear, simply because the
 Lagrangians used there imply scalar field associated particles
 with masses greater than the Planck mass ($M_{Pl}$).  In our
 approach gravity is coupled to a SU(5) grand unification
 theo\-ry (GUT), that is, to a lower energy scale.
The idea to induce gravity by a Higgs field was already discussed elsewhere
\cite{Mi77,Ze79,Ad82} and  motivation for it came to us, very similar as in
reference \cite{Ze79}, on the one hand,  from Einstein's original ideas to
incorporate the Mach principle
to GR, by which the mass of a particle should be originated from
 the interaction with all the particles of the universe, whereby
 the interaction should be the gravitational one since it couples
 to all particles, {\it i.e.},
to their masses or energies.  To realize a stronger relationship
 of the universe's particles Brans and Dicke \cite{BrDi61} introduced
their scalar-tensor-theory of gravity, letting the
 active as well as the passive gravitational mass,
that is Newton's gravitational "constant" , be  a scalar function
determinated by the distribution of particles of the universe.

On the other hand, in modern particle physics the inertial mass is
 generated by the interaction with the Higgs field, and it is
 emphasized that the successful
Higgs mechanism also lies precisely in the direction of Einstein's idea of
producing mass by a gravitational-like  interaction.  One can
show \cite{DeFrGh90,DeFr91,Fr92} that the Higgs field as  source of
the inertial mass of the elementary particles mediates a scalar
 gravitational interaction, however of Yukawa type, between those
 particles which become massive as a
consequence of the spontaneous symmetry breaking (SSB):  The
 masses are the source of the scalar Higgs field and the
Higgs field acts back by its gradient on the masses in the momentum law.

Then, due to the equivalence principle, it seems natural to identify both
approaches.  For this reason, it was proposed \cite{DeFrGh92} a scalar-tensor
theory of gravity where the isospinvalued Higgs field of elementary particles
plays also simultaneously the role of a variable gravitational constant
instead of the scalar field introduced by Brans and Dicke.

In this paper, we discuss some cosmological consequences
of this theory of gravity coupled to an isotensorial Higgs field,
which breaks down to give rise to both the X-, Y-boson masses and
Newton's gravitational constant.   Since the
symmetry breaking process of the SU(5) model could be expected to
occur in the physical universe, we are
considering inflation there.

Our study carries out, at least, two types of
inflationary models: a modified version of
"new" and "chaotic" inflation, depending
essentially on the initial conditions that
the universe chooses for the Higgs field
at the beginning of time.  An interesting
feature of the models here is that
$\lambda$ fine-tuning is not necessary, just
because of the given natural relationship of
the different mass scales of particle physics.

\section{INDUCED GRAVITY IN THE SU(5) GUT}

The scalar-tensor-theory with Higgs mechanism is based on the
Lagrange-density \cite{Fr92,DeFrGh92} with units $\hbar = c = k_{B} = 1$ and
the signature (+,-,-,-) :
\begin{equation}
\label{eql}
{\cal L}= \left[ \frac{\alpha}{16 \pi} \mbox{$ tr \Phi^{\dag} \Phi$} \,\ R
+ \frac{1}{2}
 tr \Phi^{\dag}_{|| \mu} \Phi^{|| \mu}
- V(\mbox{$ tr \Phi^{\dag} \Phi$}) +  L_{M} \right] \sqrt{- g}
\end{equation}
where $R$ is the Ricci scalar, $\Phi$ is the SU(5) isotensorial
Higgs field, the symbol $|| \mu$ means in the following the
 covariant derivative with respect to all gauged groups
and represents in (1) the  covariant gauge derivative: $\Phi_{||} =
\Phi_ {| \mu } + ig [A_ \mu , \Phi]$ where
$A_{\mu} = A_ \mu {}^a \tau _a $ are
the gauge fields of the inner symmetry group, $\tau_a$  are its generators
and ${\rm g}$ is the coupling constant of the gauge group
($|\mu $ means usual partial derivative); $\alpha $ is a
dimensionless parameter to regulate the strenght of  gravitation
and $ L_{M}$ contains the fermionic and massless bosonic fields, which
belong to the inner gauge-group SU(5); $V$ is the Higgs potential.

Naturally from the first term of equation (\ref{eql}) it follows
that $\alpha~\mbox{$ tr \Phi^{\dag} \Phi$}$ plays the role of a
variable reciprocal gravitational "constant".   The aim of
 our theory is to obtain GR as a final effect of a symmetry
 breaking process and on that way to have  Newton's gra\-vi\-ta\-tional
 constant $G$ induced by the Higgs field; similar theories have
 been  considered
 to explain  Newton's gravitational constant in the context of
 spontaneous
 symmetry breaking process to unify gravity with other fields
 involved in
 matter interactions,  see refs. \cite{Mi77,Ze79,Ad82} .

In  the minimal SU(5) GUT the Higgs field which breaks the $SU(5)$ symmetry
to $SU(3)_{C} \times SU(2)_{W} \times U(1)_{HC}$, is in the adjoint
representation a 5x5 traceless matrix taking the form in the unitary gauge
\begin{equation}
\label{eqma}
\Phi = \phi {\bf N} ~~,~~{\bf N} \equiv ~ \sqrt{2/15} ~diag(1,1,1,-3/2,-3/2) ,
\end{equation}
($\phi$ real valued function).  In this paper we analyze the cosmological
consequences when this symmetry breaking is responsible for the
generation of gravitational constant as well
as the $SU(5)$ standard particle content, {\it i.e.}, the X-, Y-boson masses.

The  Higgs-potential takes the form, using equation (\ref{eqma}),
\begin{equation}
\label{eqph}
V(\mbox{$ tr \Phi^{\dag} \Phi$}) = \frac{\mu ^2}{2}
\mbox{$ tr \Phi^{\dag} \Phi$} + \frac{\lambda }{4!}
(\mbox{$ tr \Phi^{\dag} \Phi$})^2 +
\frac{3}{2} \frac{\mu ^4}{\lambda } = V(\phi) = \frac{\lambda}{24}
\left( \phi^{2} + 6 \frac{\mu^2}{\lambda} \right)^2
\end{equation}
where we added a constant term to prevent a cosmological
constant after the breaking.  The Higgs ground state, $v$ , is given by
\begin{equation}
 v^2   \,\ = \,\ - \frac{6 \mu^2}{\lambda}
\end{equation}
with $V(v) = 0~$, where $\lambda$ is a dimensionless
real constant, whereas  $\mu^2~( < 0 )$ is
so far the only dimensional real constant of the Lagrangian.

In such a theory, the potential $V(\phi)$ will play the role of
a cosmological "function" (instead of a constant) during the
 period in which $\Phi$ goes from its initial value
$\Phi_o$ to its ground state $v {\bf N}$, where furthermore
\begin{equation}
\label{eqg}
G \,\ = \,\  \frac{1}{\alpha v^2}
\end{equation}
is the gravitational constant to realize from (\ref{eql}) the theory of GR
\cite{DeFrGh92}.  In this way, Newton's gravitational
 constant
is related in a natural form to the mass of the gauge bosons, that
for the case of the SU(5) GUT is,
\begin{equation}
\label{eqmx}
M_X \,\ = \,\ M_Y \,\ = \,\ \sqrt{\frac{5 \pi}{3}} {\rm g} v ~~~~~.
\end{equation}
As a consequence of (\ref{eqg}) and (\ref{eqmx}) one has that the
strength parameter for gravity, $\alpha$ , is determinated by
\begin{equation}
\alpha \,\ = \,\  \frac{10 \pi}{3} \left( {\rm g}
\frac{M_{Pl}}{ M_X} \right)^2
\end{equation}
where $M_{Pl}=1/\sqrt{2G}$ is the Planck mass and ${\rm g^2} \approx 0.02$.
In order to be in accordance with proton decay experiments,
$\alpha {}^{<}_{\sim} 10^{7}$ must be valid, since
the X-boson mass cannot be smaller than approximately $10^{15} GeV$.  For the
detailed calculations in sections 3 and 4 we take for $\alpha$ the upper limit
($\alpha = 10^{7}$).   In this way,  the coupling between
the Higgs field and gravitation is very strong:
the fact that $\alpha>>1$ is the price paid in recovering  Newton's
gravitational constant at that energy scale.  On the
other hand, for the BDT the value of
its corresponding $\alpha ~(=2 \pi / \omega )$ must be $\alpha < 10^{-2}$
to fit well the theory with the experimental data \cite{Wi93}.  Therefore,
in this respect there are important differences between our presentation
 here and the Brans-Dicke one and also with regard to most of the
induced gravity approaches, where to achieve successful inflation typically
 $\alpha << 1$ \cite{AcZoTu85,LuMaPo86}, and in that way, the existence
of a very massive particle ($> M_{Pl}$) is necessary, which after inflation
should decay into gravitons making difficult later an acceptable
nucleosynthesis scenario \cite{BaSe89,BaSe90}.

\bigskip
{}From (\ref{eql}) one calculates immediately the gravity equations of the
theory
\begin{eqnarray}
\label{eqrp}
\lefteqn{ R_{\mu \nu} -\frac{1}{2}R g_{\mu \nu} +
 {8 \pi V(\mbox{$ tr \Phi^{\dag} \Phi$})\over
\mbox{$\alpha~tr \Phi^{\dag} \Phi$}}  g_{\mu \nu} \,\ = \,\ }
\nonumber\\ &&- {8 \pi  \over \mbox{$\alpha~tr \Phi^{\dag} \Phi$}} T_{\mu \nu}
 - {8 \pi \over \mbox{$\alpha~tr \Phi^{\dag} \Phi$}} \left[
 tr \Phi^{\dag}_{( || \mu} \Phi_{|| \nu )}   -
tr  \frac{1}{2} \Phi^{\dag}_{|| \lambda} \Phi^{|| \lambda} \,\ g_{\mu \nu}
\right]
 \nonumber\\ &&
- \frac{1}{\mbox{$ tr \Phi^{\dag} \Phi$}} \left[ tr (\Phi^{\dag} \Phi)_{|
\mu || \nu} -
tr (\Phi^{\dag} \Phi)^{| \lambda}_{~~|| \lambda} \,\ g_{\mu \nu} \right]
\end{eqnarray}
where $ T_{\mu \nu} $ is the energy-momentum tensor belonging to
 $L_M \sqrt g $ in (1)
alone, and the Higgs field equations
\begin{equation}
\label{eqh}
 \Phi^{|| \lambda }{}_{|| \lambda} + \frac{\delta V}{\delta \Phi^{\dagger} } -
\frac{\alpha}{8 \pi} R \Phi \,\ = \,\
2 \frac{\delta L_{M} } {\delta \Phi^{\dag}}  ~~~~~~.
\end{equation}
Now we introduce the new real valued scalar variable
\begin{equation}
\chi \equiv \frac{1}{2} \left( \frac{\mbox{$ tr \Phi^{\dag} \Phi$}}{v^2} - 1
\right)
\end{equation}
which describes the excited Higgs field around its ground
state; for instance
$\Phi=0$ implies $\chi = -1/2$ and
$\Phi=v {\bf N} $ implies $\chi = 0$.  With this new Higgs variable
equations (\ref{eqrp}) and (\ref{eqh}) are now:
\begin{eqnarray}
\label{eqrx}
\lefteqn{ R_{\mu \nu} -\frac{1}{2}R g_{\mu \nu} +
\left[ {8 \pi  \over \alpha v^2} {V(\chi) \over (\mbox{$ 1+2 \chi$}) } \right]
g_{\mu \nu}
\,\ = \,\ }\nonumber\\ && - {8 \pi  \over \alpha v^2 }
\frac{1}{( \mbox{$ 1+2 \chi$} )} \, \hat{T}_{\mu \nu}
 - {8 \pi \over \alpha}  \frac{1}{(\mbox{$ 1+2 \chi$})^2}\left[
\chi_{ |\mu} \chi_{| \nu}  - \frac{1}{2} \chi_{ |\lambda} \chi^{|\lambda}
 g_{\mu \nu} \right] \nonumber\\ &&
- \frac{2}{\mbox{$ 1+2 \chi$}} \left[ \chi_{| \mu || \nu} - \chi^{| \lambda}
_{~|| \lambda}
g_{\mu \nu} \right]
\end{eqnarray}
and
\begin{eqnarray}
\label{eqhx}
\chi^{| \mu}_{~~|| \mu} + \frac{1}{\mbox{$(1+ \frac{4 \pi}{3\alpha})$}}
\frac{4 \pi}{3 \alpha v^2}
\frac{\delta V}{\delta \chi}
&  =  & \frac{1}{\mbox{$(1+ \frac{4 \pi}{3\alpha})$}}
 {4 \pi  \over 3 \alpha v^2} \, \hat T
\end{eqnarray}
where  $\, \hat{T}_{\mu \nu }$ is  the {\it effective} energy-momentum tensor
 given by
\begin{equation}
\label{eqtt}
\, \hat{T}_{\mu \nu} \,\ = \,\ T_{\mu \nu} +
\frac{(\mbox{$ 1+2 \chi$})}{4 \pi}
\left( A^{a}_{~\mu} A^{b}_{\nu} - \frac{1}{2} g_{\mu \nu}
A^{a}_{~\lambda} A^{b \lambda}   \right) ,
\end{equation}
where $M^{2}_{ab}$ is the gauge boson mass square matrix.

The continuity equation
(energy-momentum conservation law) reads
\begin{equation}
\label{eqct}
\, \, \hat T^{~\nu}_{\mu~~ || \nu} \,\ = \,\ 0 ,
\end{equation}
which has no source since in the present theory, SU(5) GUT, all the fermions
remain massless after the first symmetry breaking and no baryonic matter
is originated in this way.

Another important difference with the BDT is due to the existence
of the potential term, which shall play the role of a positive
 cosmological
function (see the bracket on the left hand side of equation
 (\ref{eqrx})); it takes in terms of $\chi$ the simple form,
\begin{equation}
\label{eqpot}
V(\chi)  \,\ = \,\  {\lambda v^4 \over 6} \chi^2
\end{equation}
which at the ground state vanish,
$ V(\chi =0 )  =   0  $.
{}From equation (\ref{eqrx}) and (\ref{eqg}) one
recovers GR for the ground state
\begin{equation}
 R_{\mu \nu} -\frac{1}{2}R g_{\mu \nu} \,\ = \,\ - 8 \pi G ~ \, \hat{T}_{\mu
\nu}
\end{equation}
with the effective energy-momentum tensor (\ref{eqtt}).  Newton's
gravitational ''function'' is $G(\chi)= \frac{1}{\alpha v^2}\frac{1}{
\mbox{$ 1+2 \chi$}}$
and Newton's gravitational constant
$G(\chi=0)= G $.
\bigskip

{}From (\ref{eqhx}) one can read directly, if the explicit form of $V(\chi)$ is
introduced, the mass of the Higgs boson
$M_{H}$ , and therefore its Compton range $l_{H}$,
\begin{equation}
\label{eqm}
M_{H} \,\ = \,\  \sqrt{\frac{\frac{4 \pi}{9 \alpha}\lambda v^2}{\mbox{$(1+
\frac{4 \pi}{3\alpha})$}}}
{}~,~~ l_{H} \,\ = \,\ \frac{1}{M_{H} } ~~,
\end{equation}
that is, the $\chi-$field possesses a finite range, a characteristic which
accounts for the di\-ffe\-ren\-ce between this theory and the BDT. {}From
equation  (\ref{eqm})  one sees that the mass of
the Higgs particle is a factor $\sqrt{\frac{4 \pi}{3 \alpha}} \approx 10^{-4}$
smaller than the one derived from the SU(5) GUT
without gravitation.  This is a very interesting property since
 the Higgs mass determines the scale of the symmetry breaking, or
 equivalently, $ \sqrt{\lambda}/\alpha$ will be in a
natural way a very small value, avoiding a fine-tuning of
$\lambda$ in order to accomplish a successful universe (see next sections).

Next, we proceed to investigate the cosmological consequences of such
a theory.

              \section{FRW- MODELS}

Let's consider a Friedman-Robertson-Walker (FRW) metric.  One has
with the use of (\ref{eqg}) that equations (\ref{eqrx}) are
reduced to:

\begin{eqnarray}
\label{eqap}
 \frac{\mbox{$\dot{a}$}^2 + \epsilon }{a^2} \,\ = \,\
\frac{1}{\mbox{$ 1+2 \chi$}} \left(
{8 \pi G  \over 3 }  [ \rho  + V(\chi)  ]
- 2 \frac{\mbox{$\dot{a}$}}{a} \mbox{$\dot{\chi}$}+  {4 \pi  \over 3 \alpha}
\frac{\mbox{$\dot{\chi}$}^2}{\mbox{$ 1+2 \chi$}} \right)
\end{eqnarray}
and
\begin{eqnarray}
\label{eqapp2}
 \frac{\mbox{$\ddot{a}$}}{a}  =
\frac{1}{\mbox{$ 1+2 \chi$}} \left( {4 \pi G  \over 3 }   [ -\rho  -3 p +2
V(\chi) ]
 -  \mbox{$\ddot{\chi}$}- \frac{\mbox{$\dot{a}$}}{a} \mbox{$\dot{\chi}$} -
\frac{8 \pi}{3 \alpha}  \frac{\mbox{$\dot{\chi}$}^2}{\mbox{$ 1+2 \chi$}}
\right)
\end{eqnarray}
where $a=a(t)$ is the scale factor, $\epsilon $ the curvature constant
($\epsilon = 0,~+1 ~or~ -1$ for a flat, closed or open space, correspondingly),
  $\rho$ and $p$ are the matter density and pressure assuming that the
effective  energy momentum tensor (\ref{eqtt}) has in the classical
limit the structure of that of a perfect fluid.  An overdot stands for a time
derivative.

In the same way equation (\ref{eqhx}) results in:
\begin{eqnarray}
\label{eqxp}
\mbox{$\ddot{\chi}$}+ 3 \frac{\mbox{$\dot{a}$}}{a} \mbox{$\dot{\chi}$}+
M_{H}^{2} \chi \,\ = \,\
\frac{4 \pi G}{3} \frac{(\rho - 3p)}{\mbox{$(1+ \frac{4 \pi}{3\alpha})$}}
\,\ ,
\end{eqnarray}
where the Higgs potential is already inserted, {\it i.e.}
\begin{equation}
V(\chi)={\lambda v^4 \over 6} \chi^2 =
\mbox{$(1+ \frac{4 \pi}{3\alpha})$}\frac{3}{8 \pi G } M_{H}^{2} \chi^2  ~~~~~.
\end{equation}
The Higgs mass demarcates
the time epoch for the rolling over of the potential, and therefore
for inflation.  Note that $V(\chi) \sim  M_{Pl}^{2}  M_{H}^{2}\chi^2$ ; this
fact is due to the relationship (\ref{eqg}) to obtain GR once
the symmetry breaking takes place.

The continuity equation (\ref{eqct}) takes the simple form
\begin{equation}
\label{eqc}
\dot{\rho} + 3 \frac{\mbox{$\dot{a}$}}{a}  (\rho + p)  \,\ = \,\ 0
\end{equation}
which is sourceless, meaning that the Higgs mechanism produces no entropy
processes,  although the Higgs field is coupled
to the perfect fluid through equation (\ref{eqxp}).
If one takes the equation of state of a barotropic fluid, {\it i.e.}
$p = \nu \rho$ with the dimensionless constant $\nu$,
equation (\ref{eqc}) can be easily integrated
$ \rho  =  \frac{M}{a^{3 (1+\nu)}} $,
where $M$ is the integration constant.

At this point we would like to make several remarks concerning
the  scale factor
equations (\ref{eqap}) and (\ref{eqapp2}).  First of all, the equations
allow static solutions
( $ \mbox{$\dot{\mbox{$\dot{a}$}}$}= \mbox{$\ddot{a}$}= \mbox{$\dot{\chi}$}=
\mbox{$\ddot{\chi}$}= \dot{\rho} \equiv 0 $) for
dust particles ($\nu =0$) :
\begin{eqnarray}
\label{eqss}
\chi = 1 ~~,~~
a^2= \frac{1}{\mbox{$(1+ \frac{4 \pi}{3\alpha})$}} \frac{1}{M_{H}^{2}}
{}~ and~~ \rho= \frac{3}{2 \pi} \mbox{$(1+ \frac{4 \pi}{3\alpha})$}M_{Pl}^{2}
M_{H}^{2} .
\end{eqnarray}
For the radiation case, $\nu = 1/3$, there are no static solutions.

For the dynamical behaviour one notes that the Higgs potential is indeed
a positive cosmological function, which
corresponds to a positive mass density and a negative pressure
(see equations (\ref{eqap}) and (\ref{eqapp2})), and represents an ideal
ingredient to
have inflation.  But, on the other hand, there is a negative
contribution to the acceleration equation (\ref{eqapp2})
due to the Higgs-kinematic terms, {\it i.e.} terms involving
$\mbox{$\dot{\chi}$}$ and
$\mbox{$\ddot{\chi}$}$ ;  terms involving the factor $1/ \alpha \sim 10^{-7}$
are simply too small compared to the others and can be neglected.

For inflation it is usually taken that $\mbox{$\ddot{\chi}$}\approx 0$, but
in fact
the dynamics should show up this behaviour or at least certain
consistency.  For instance, in GR with the {\it ad hoc}
inclusion of a scalar field $\phi$ as a source for the inflation, one
has that at the "slow rollover" epoch $\ddot{\phi} \approx 0$  and
therefore $\dot{\phi} = - V^{\prime}/3H$, which implies that

\begin{equation}
\label{eqrollp}
\frac{\ddot{\phi}}{3H \dot{\phi}} =
- \frac{V^{\prime\prime}}{9 H^2} + \frac{1}{48 \pi G}
\left(\frac{V^{\prime}}{V}\right)^2 << 1 ~~~,
\end{equation}
where $H = \mbox{$\dot{a}$}/a$ is the Hubble expansion rate (a prime denotes
the derivative
with respect to the corresponding scalar field, see
reference \cite{StTu84}).  In the present theory, if one considers
the Higgs potential term in equation (\ref{eqap}) as the dominant one
\footnote{{}From now on, we shall always consider the dynamics to be dominated
by the Higgs terms, in eqs. (\ref{eqap})-(\ref{eqxp}),
 instead of the matter density term, from which it is not possible to drive
 inflation.} and equation (\ref{eqxp}) without source,
 {\it i.e.} $p = \frac{1}{3} \rho$, one has indeed an extra
 term due to the variation of Newton's "function" $G(\chi)$, that is

\begin{equation}
\label{eqrollx}
\frac{\ddot{\chi}}{3H \dot{\chi}} =
\left( \frac{1}{1+\frac{4 \pi}{3 \alpha}}\right)\frac{4 \pi G}{3 }
 \left[- \frac{V^{\prime\prime}}{9 H^2} + \frac{1}{48 \pi G}
 \left(\frac{V^{\prime}}{V}\right)^2 (\mbox{$ 1+2 \chi$}) - \frac{1}{24 \pi G}
 \frac{V^{\prime}}{V} \right] ~~~ .
\end{equation}
If $\chi<0$ the last term does not approach to zero during the rolling
down process for $\alpha >> 1$.  Thus,
one has instead of a "slow", rather a "fast" rollover dynamics of the
Higgs field along its potential down hill.  On the other side, for
$\chi >> 1$ there is indeed a "slow" rollover dynamics.

With this in mind one has to look carefully at the contribution of
$\mbox{$\ddot{\chi}$}$ : if one brings $\mbox{$\ddot{\chi}$}$ from
(\ref{eqxp}) into (\ref{eqapp2}) one
has that $M_{H}^{2} \chi$  competes with the potential
term $M_{H}^{2} \chi^2$, and during the rolling down of the potential, when
$\chi$ goes from  $-1/2$ to $0$, $M_{H}^{2} \chi < 0$  dominates the
dynamics, and therefore instead of inflation one ends with
deflation or at least with a contraction era for the scale factor.

Resuming, if one starts the universe evolution with an ordinary new inflation
scenario ($\chi_{o} < 0$ \footnote{The subindex "$o$" stands for the initial
values (at $t=0$)  of the corresponding variables.}), it implies in
this theory a "short" deflation instead of a "long"
inflation period, since the Higgs field goes relatively fast to
 its minimum.  This feature should be present in theories
of induced gravity with $\alpha > 1$ and also for the BDT with
this type of potential (see for example the
field equations in reference \cite{Wa92}).  Considering
 the opposite limit, $\alpha<1$,
induced gravity models \cite{AcZoTu85} have proved to be successful
for inflation,
 also if one includes others fields \cite{AcTr89}; induced gravity theories
 with a Coleman-Weinberg potential are also shown to be treatable for
 a very small coupling constant $\lambda$ with $\chi_o<0$
 \cite{LuMaPo86,Po86}, or with $\chi_o>0$ \cite{Po85}
and $\alpha < 1$ as well as $\alpha > 1$ \cite{Sp84}.  For extended or
hyperextended inflation
models \cite{LaSt89,StAc90} this problem doesn't arise due to the
presence of vacuum energy during the rollover stage of evolution,
 which is
supposed to be greater than the normal scalar field contribution.

With this concern one has to prepare a convenient scenario for the
Universe to begin with.  In the next section we analyse the
initial conditions of our models.
\section{INITIAL CONDITIONS AND INFLATION}

The initial conditions that we have chosen are simply
$\mbox{$\dot{a}$}_o = \mbox{$\dot{\chi}$}_o =0 $.  Equations (\ref{eqap})
to (\ref{eqxp}) must
satisfy the following relations:

The size of the initial universe is if $\epsilon =1$
\begin{eqnarray}
\label{eqao}
a^2_o = \frac{  \mbox{$ 1+2 \chi_o$} }{
  {8 \pi G  \over 3} \rho_o + \mbox{$(1+ \frac{4 \pi}{3\alpha})$}M_{H}^{2}
\chi_o^2 } ,
\end{eqnarray}
its acceleration has the value
\begin{eqnarray}
\label{eqappo}
\frac{\mbox{$\ddot{a}$}_o}{a_o}   =  \frac{1}{\mbox{$ 1+2 \chi_o$}}
 \Bigg\{ - {4 \pi G  \over 3 } \left[  1+ \frac{1}{\mbox{$(1+ \frac{4 \pi}
{3\alpha})$}}
 + \left( 1- \frac{1}{\mbox{$(1+ \frac{4 \pi}{3\alpha})$}} \right) 3 \nu
\right] \rho_o \nonumber\\
 + \mbox{$(1+ \frac{4 \pi}{3\alpha})$}M_{H}^{2} \chi_o^2 + M_{H}^{2} \chi_o
\Bigg\}
\end{eqnarray}
and for the Higgs field it is valid
\begin{eqnarray}
\label{eqbx}
\mbox{$\ddot{\chi}$}_o  + M_{H}^{2} \chi_o \,\ = \,\
\frac{4 \pi G}{3} \frac{(1-3\nu)}{\mbox{$(1+ \frac{4 \pi}{3\alpha})$}} \rho_o .
\end{eqnarray}
The initial values $\rho_o$ and $\chi_o$ as well as  $M_{H}$
are the cosmological parameters to determine the  initial
conditions of the Universe.  The value of $M_{H}$ fixs
 the time scale for which the Higgs field breaks down into its ground
 state.   In order to consider  the Higgs-terms as the dominant ones
 (see footnote 1), one must choose the initial matter density
$\rho_{o} < \frac{3 \chi^{2}_{o}}{4 \pi} \mbox{$(1+ \frac{4 \pi}{3\alpha})$}
M_{Pl}^{2} M_{H}^{2}$.
Let's say,  for a Higgs mass
$M_H \sim 10^{-1} M_{X}= 10^{14} GeV$, one has a typical time of
$M_{H}^{-1} \sim 10^{-38}$ sec, and therefore
$\rho_{o} < \frac{3 \chi^{2}_{o}}{4 \pi} 10^{66} GeV^{4} $.

\smallskip

The question of the choice of the initial value $\chi_o$ is open: for
example, for "new inflation" $\chi_o < 0$ \cite{Li82,AlSt82,Sta82}, whereas
for "chaotic inflation" \cite{Li83} $\chi_o > 0$.  Therefore, we are
considering both cases, which imply two different cosmological scenarios.

{\bf Scenario (a) ($\chi_{o}<0$):} {}From equation (\ref{eqao}) it follows
that if the initial value of the
Higgs field is strictly $\chi_{o} = -1/2$, the universe
possesses a singularity.
If the Higgs field sits near to its metastable equilibrium point at
the beginning ($\chi_o {}^{>}_{\sim} -1/2$, $\Phi_o {}^{>}_{\sim} 0$),
than $\chi$ grows up since $\mbox{$\ddot{\chi}$}_o > 0$,
and from equation (\ref{eqappo}) one gets
 that $\mbox{$\ddot{a}$}_o < 0$ , {\it i.e.} a maximum point for $a_o$;
thus at the beginning
 one has a contraction instead of an expansion.   Let's call this
{\it rollover contraction}, see figure 1(a).

Normally it is argued that in BDT with a constant (or slowly varying)
potential producing a finite vacuum energy
density,  the  vacuum energy is dominant and is used to both to
expand the universe and to increase the value of the scalar field.  This
"shearing" of the vacuum energy to both pursuits  is
the cause of a moderate power law inflation instead of an
exponential one \cite{KaKaOl90}.  In this scenario the universe begins
with a contraction, and therefore the same shearing mechanism, moreover
here due to the Higgs field, drives a "friction" process
for the contraction, due to the varying of $G(\chi)$,
 making the deflation era always weaker.
Furthermore, one can  see from equation (\ref{eqappo}) that the cause
of the deacceleration in scenario (a) is the negative value of $\chi$;
then if $\mbox{$\dot{a}$}< 0$  from equation  (\ref{eqxp}) it
follows $\mbox{$\ddot{\chi}$}\sim - \frac{\mbox{$\dot{a}$}}{a}
\mbox{$\dot{\chi}$}>0$, which implies
an "anti-friction" for $\chi$ that tends to reduce the
contraction, see ref. \cite{CaPo94}.

One may wonder if the rollover
contraction can be stopped.   As long as $\chi$ is
negative the contraction will not end, but if $\chi$ goes to positive
values, impulsed by special initial conditions, one could eventually have
that the dynamics dominating term, $ M_{H}^{2} (\chi^2 + \chi)$, be positive
enough to drive an expansion.  But due to the nature of equation
(\ref{eqxp}), if $\chi$ grows up , the term
$ M_{H}^{2} \chi $ will bring it back to negative values and
causes an oscillating behaviour around zero, its equilibrium state,
with an amplitude which is damped
with time due to the redshift factor $3H \mbox{$\dot{\chi}$}$.
Therefore, one has to seek special values of $\chi_o$ ,
which will bring $\chi$ dynamically from negative values to great enough
positive values to end up with sufficient e-folds of inflation.  This
feature makes clear that this scenario is not generic for inflation, but
depends strongly on special initial conditions; in this sense, this is
another type of fine-tuning, which is however always present by choosing the
initial value of the inflaton field.  For instance, in the SU(5) GUT
energy scale with $M _H = 10^{14}$ GeV, one finds by numerical integration
that  obtaining the required inflation implies  $\chi_o \approx -0.15509$
(see figure 2(a) and 3) but not a very different number than this,
otherwise the deflation era doesn't stop and the universe evolves to an
Einstein universe with a singularity; one could ask
 whether GR singularities  are an inevitable consequence of
 particle physics.  It was also assumed, of course, that during
 the deflation phase the stress energy of other fields, e.g.
 radiation or non-relativistic matter, are smaller than the Higgs one,
 otherwise an expansion follows.

It is interesting to note that in the SU(5) GUT with a finite-temperature
effective potential coupled conventionally to GR, inflation takes place only
after the temperature $T {}^<_{=} 0.05 \sigma$ ($\sigma=2$x$10^{15} GeV$) and
$| \phi - \sigma | > 0.03 \sigma$, since there the "inflation pressure"
$\rho(\phi,T)+ 3 p(\phi,T)$  is negative.   Then if these requirements are not
fulfilled, there can appear a deflation era in the universe, see
\cite{CaPo94}.  In our presentation, temperature co\-rrections are not
considered, but
the only possible initial Higgs values for a successful scenario (a) are
($\chi_{o} \sim -0.1$ or $\phi \sim 0.8 v$) those which
correspond to the beginning of inflation in GR with temperature corrections.
In both cases inflation begins when the Higgs initial value  is not very far
from its equilibrium state $v$, that means, $\phi$ is not located very close
to zero as usual. Furthermore, if we were to consider temperature corrections,
a negative contribution to the potential comes in, see ref. \cite{CaPo94},
making the deflation stronger, since it should be
added to the already considered coming from $\mbox{$\ddot{\chi}$}>0$ in
equation (\ref{eqapp2}).  However,
for the above mentioned special values  of the Higgs field, because of the
 temperature co\-rrections is now expected a positive contribution to
the acceleration of
the scale factor, which should compete with  $\mbox{$\ddot{\chi}$}$ to
determine the scale factor evolution.  Finally, only
for some special values of $\chi_o$, a short deflation
is followed by a successful inflation period.

{\bf Scenario (b) ($\chi_{o}>0$):}  One could  think
about initial conditions whereby
the Higgs terms $ \chi^2_o + \chi_o > 0$ dominate the dynamics to
have a minimum for $a_o$, {\it i.e.} $\mbox{$\ddot{a}$}_o > 0$,
and to begin on "a right
way" with expansion {\it i.e.} inflation.  That means one should start with
a value  $\chi_o>0$ (far from its  minimum) positive enough to
render sufficient e-folds of inflation.  Thus, the "effective" inflation
potential part is similar to the one proposed in the "chaotic" inflationary
model \cite{Li83} due to the form of the potential and to the
value of $\chi_o > 0$ to have the
desired inflation.  In other words both
this and chaotic inflation scenarios
are generic \cite{GoPi92}, see figure 1(b).

Now let's see how the dynamics of both models works: The curvature term
$\epsilon/a^{2}$ in
(\ref{eqap}) can be neglected only after inflation began; during
the rollover contraction (a) it plays an important
role. The terms $\frac{\dot{a}}{a} \mbox{$\dot{\chi}$}$  will be comparable to
$8 \pi G V(\chi)/3 $
till the high oscillation period
($H < M_{H}$) starts.  For instance, in
the chaotic scenario (b),   the slow rollover condition
$\mbox{$\ddot{\chi}$}\approx 0$ \footnote{this condition is equivalent to
both known prerequisites
$3 H >> \ddot{\phi}/\dot{\phi},~ 3 \dot{\phi}/\phi$ of induced gravity.}
is valid, which implies  $\mbox{$\dot{\chi}$}/ \chi = - M^{2}_{H}/3H$,
and then from
\begin{equation}
\label{eqtc}
H^{2} \approx \frac{1}{\mbox{$ 1+2 \chi$}} \left[  M^{2}_{H} \chi^{2} - 2
H \mbox{$\dot{\chi}$}\right]  ~~
\end{equation}
(with $\mbox{$\dot{\chi}$}<0$)  it follows that for $\chi > 2/3$  the
Hubble parameter
will be dominated by the potential term to have:
\begin{equation}
\label{eqhh}
H \approx M_{H} \frac{\chi}{\sqrt{\mbox{$ 1+2 \chi$}}}   ~~ ,
\end{equation}
which for $\chi >>1$ goes over into $H/M_{H} \sim \sqrt{\chi /2}>>1$ giving
cause for the slow rollover ("chaotic") dynamics.  Indeed, the rollover time is
$\tau_{roll} \sim 3 H/M^{2}_{H} $ , {\it i.e.}

\begin{equation}
\label{eqroll}
H \tau_{roll} \,\ \approx \,\ 3 H^{2}/M^{2}_{H} \,\ \approx \,\
 3 \frac{\chi^{2}}{\mbox{$ 1+2 \chi$}} {}^{>}_{\sim}  65 ~~ ,
\end{equation}
yielding  enough inflation; for $\chi_{o} >>1 $ it follows
$H \tau_{roll}  \sim 3 \chi_{o}/2 {}^{>}_{\sim} 65$, which implies
$ \chi_{o} {}^{>}_{\sim} 130/3$ ($\phi_o {}^{>}_{\sim} 9.3 v $); this value
can be checked by
numerical integration, see figure 2(b) and 4. Note that $H/M_H$ doesn't
depend on the energy scale of inflation but on the initial
 value $\chi _o$, in other words, enough inflation is performed
automatically and independently of $\alpha$, as was pointed out in
ref. \cite{FaUn90}. Then if one considers inflation at a lower
 energy scale, $\chi _o$ can be smaller than 130/3, since at that
 smaller energy scale the e-folds required are less.

On the other side if $\chi_{o}$  is negative,   the rollover
contraction phase in scenario (a) happens, but in this case equation
(\ref{eqtc})  indicates $ H/M_{H} \approx |\chi| < 1$, that is, the
scale factor evolves slower than the Higgs field; and with
$\chi_o \approx -0.15509 $,  $\chi $ evolves to values greater or equal than
$ 130/3 $ to gain conditions very similar to scenario (b), see figures 3 and 4.

Summarizing, for the two possibilities of universe's models, one
has the following: In the chaotic scenario (b), the initial value should be
$\chi_{o} > 130/3$ in order to achieve sufficient e-folds of inflation.
 And in
the scenario (a), only for special initial values of the Higgs field
($\chi_o \approx  -0.15509$) , the universe undergoes a small contraction
which goes over automatically into a sufficiently long inflation period;
otherwise, for other initial negatives values of $\chi_{o}$, the universe
contracts to a singularity.

At the end of inflation the Higgs field begins to oscillate with a frequency
$M_{H} > H$ and the universe is now  dominated by the
oscillations, which drive a normal Friedmann regime \cite{Tu83}.  This can
be seen as follows:  First when  $H {}^{>}_{\sim} M_{H}$   with
$H \approx const.$, $~ \chi
\approx e^{-3H/2~t} cos M_{H} t$ is valid, later on when
$H << M_{H}$, $H\sim 1/t$ and $ \chi \sim 1/t ~ cos M_{H} t $ give rise to
$a \sim t^{2/3}$, {\it i.e.} a matter dominated universe with  coherent
oscillations, which will hold on if the Higgs bosons
do not decay; in figures 5 and 6  the
behaviour of the scale factor and the Higgs field is shown until the time
$100 M^{-1}_{H}$; the numerics fit very well the  "dark"
 matter dominated solutions.  Let us consider this possibility more in
detail: then, the average over one oscillation of the absolute value of the
effective energy density of these oscillations,
$\rho_{\chi} \sim V(\chi) = \frac{3}{4 \pi} M^{2}_{Pl} M^{2}_{H} \chi^{2}$,
is such that

\begin{equation}
\label{eqrx*}
\frac{\rho_{\chi}}{\rho_{\chi_{*}}} = \left( \frac{t_{*}}{t} \right)^2
\end{equation}
where $t_{*}$ is the time when the rapid oscillation regime begins.
If  $M_{H} = 10^{14} GeV$, $t_{*} \approx 10^{2} M^{-1}_{H} = 10^{-36} sec$
for which $\chi_{*} \approx 10^{-3}$, then  it follows that
$\rho_{\chi_{*}} = 10^{60} GeV^{4}$. Thus nowadays when
$t_{n}\sim 10^{17} sec$, one should have
\begin{equation}
\label{eqrxn}
\rho_{\chi_{n}} = 10^{60} \left( \frac{10^{-36}}{10^{17}} \right)^2 GeV^{4} =
10^{-46} GeV^{4} \approx 10^{-29} grs/cm^{3}
\end{equation}
{\it i.e.} the Higgs oscillations could solve the missing mass
 problem of cosmology, implying the existence of cold dark matter,
since after some time
as the universe expands the Higgs particles will have a very slow momentum
owing to their big mass.  But, in going so far, we have neglected other
meaningful facts
in the evolution of our physical universe, as could be the influence of
such an oscillation for baryogenesis and/or for nucleosynthesis; their
repercussion must be still investigated, nevertheless not at the scope
of this paper.

There is, however, another problem: if one tries to explain the today
observed baryonic mass of the universe, then owing to inflation this
mass is given by
$M(t) \approx M/a^{3 \nu}$  and is too small.  For
solving this problem one has to assume that some amount of the Higgs
oscillations decay into baryons and leptons.
 At the time around $t_{*}$ the Higgs field should decay into
other particles  with a decay width $\Gamma_H$ to give
place to a normal matter or radiation universe expansion, producing
the reheating of the universe \cite{AlTuWi82,DoLi82,AbFaWi82}.  If reheating
takes place, the still remaining energy of the scalar field at $t_{*}$ is
converted into its decay products.  This would mean that the cosmological
"function" disappears to give then rise to  the known matter of the universe.
But
if  coherent oscillations still stand, they are the remanents of that
cosmological "function", which is at the present  however
 invisible to us in the form of cold dark matter.
Now suppose that they really did decay.
 Mathematically, the way of stopping the oscillations or to force
 the decay is to introduce a term $\Gamma_{H} \mbox{$\dot{\chi}$}$ in equation
 (\ref{eqxp}).  The universe  should then reheat up to the
 temperature $T_{RH} \approx \sqrt{M_{Pl} \Gamma_{H}}$ where
 $\Gamma_{H}$ depends, of course, on the decay products.  For example, if the
coherent oscillations decay into two light fermions \cite{AcZoTu85}
it is valid:
\begin{equation}
\label{eqga}
\Gamma_{H} \,\ \approx  \,\ g^{2} M_{H} \,\ \approx \,\ g^{2}
\frac{2\sqrt{2 \pi}}{3} \frac{\sqrt{\lambda}}{\alpha} M_{Pl}
\end{equation}
where we have used equation (\ref{eqm})  with $\mbox{$(1+ \frac{4 \pi}
{3\alpha})$}\approx 1$.  For the
reheating this would mean that
\begin{equation}
\label{eqtr1}
T_{RH} \approx \sqrt{M_{Pl} \Gamma_{H}} =
g \sqrt{ \frac{2\sqrt{2 \pi}}{3}\frac{\sqrt{\lambda}}{\alpha}} M_{Pl}   ~~.
\end{equation}
Now consider again equation (\ref{eqm}), from which it follows that

\begin{equation}
\label{eqlambda1}
\lambda \,\  = \,\  \frac{9 \alpha}{4 \pi} \left( \frac{M_{H}}{v}
\right)^{2}  \,\  = \,\
\frac{25 \pi}{2} g^{4} \left(  \frac{M_{Pl}}{M_{X}} \frac{M_{H}}{M_{X}}
\right)^{2} .
\end{equation}
If one chooses,
$\frac{M_{Pl}}{M_{X}} = 10^{4} ,~ \frac{M_{H}}{M_{X}} = 10^{-1}$  one has that
\begin{equation}
\label{eqlambda2}
\lambda \,\  = \,\  \frac{25 \pi}{2}  10^{6} g^{4} ,
\end{equation}
for which one has not the usual $\lambda$ fine-tuning,
$\lambda < 10^{-12}$, see ref. \cite{AcZoTu85}.
Coming back to the reheat temperature, one has that
\begin{equation}
\label{eqtr2}
T_{RH} \approx g^{2} 10^{2}  \frac{M_{Pl}}{\sqrt{\alpha}}
\,\ = \,\   10^{15} GeV
\end{equation}
which should be enough for baryogenesis to occur.
The right baryon asymmetry could be generated in this model if
the masses of the Higgs triplets
(as decays products) are
between $10^{11}-10^{14} GeV$ , see ref. \cite{AbFaWi82}.
Also one should be aware of the production of gravitational
radiation  as a decay output of the oscillations \cite{BaSe89,BaSe90};
however,  it could be also possible that other decay
channels are  important, since the symmetry breaking takes
here place at a much more smaller energy scale than the Planck one.

The contrast of density perturbations $\delta\rho/ \rho$ can be considered
in scenario (b), or in scenario (a)  when $\chi$ has evolved to its positive
values to have very similar slow rollover conditions as in (b).  Then one has
\cite{FaUn90,MaSa91}
\begin{equation}
\label{eqc1}
\frac{\delta\rho}{\rho} \mid_{t_{1}} \approx
\frac{1}{\sqrt{1+\frac{3\alpha}{4 \pi}}}
H \frac{\delta\chi}
{\mbox{$\dot{\chi}$}} \mid_{t_{1}} =
\sqrt{\frac{3}{\pi \alpha}}
\frac{M_{H}}{v}\frac{\chi^{2}}{\mbox{$ 1+2 \chi$}} \mid_{t_{1}}
 \end{equation}
where $t_{1}$ is the time when the fluctuations of the scalar
field leave $H^{-1}$ during inflation.  At that time, one finds that
\begin{equation}
\label{eqc2}
\frac{\delta\rho}{\rho} \mid_{t_{1}} \approx
\frac{2}{\sqrt{3}}
\frac{\sqrt{\lambda}}{\alpha}
\frac{\chi^{2}}{\mbox{$ 1+2 \chi$}} \mid_{t_{1}}
\approx
\sqrt{\frac{3}{2 \pi}}
\frac{M_{H}}{M_{Pl}}
\frac{\chi^{2}}{\mbox{$ 1+2 \chi$}} \mid_{t_{1}} \,\ \approx \,\
 10 \frac{M_{H}}{M_{Pl}}  = 10^{-4}
 \end{equation}
which corresponds exactly to the ratio of the masses already chosen and fixs
an upper limit to the density perturbations that give rise to the
observed astronomic structures, for which $\chi (t_{1}) {}^{>}_{\sim} 30$,
corresponding to more than 50  e-folds before inflation ends.
The natural smallness of $\sqrt{\lambda }/\alpha$ avoids the
usual fine-tuning of $\lambda$ necessary
to keep $\delta\rho / \rho$ sufficient small.
The perturbations on the microwave background
temperature are for the same reason approxi\-ma\-tely well
fitted.  The gravitational wave perturbations considered
  normally should also be small \cite{AcZoTu85} :
\begin{equation}
\label{eqgw}
h_{GW} \approx \frac{H}{M_{Pl}}  \approx   \frac{M_{H}}{M_{Pl}}
\sqrt{ \frac{\chi}{2} } \,\ \sim  10^{-5}
\end{equation}
again to be evaluated when the scale in question crossed outside $H^{-1}$
during inflation.

\section{CONCLUSIONS}
The scalar-tensor-theory with Higgs mechanism applied to the SU(5) GUT
can drive successful inflation without forcing the parameter $\lambda $ of
particle physics to be very small.  This is performed by means of
 the natural relationship
 among the fundamental masses of physics at that energy scale: The
 Higgs-, X-boson- and Planck-masses, on the strength of equations
 (\ref{eqg}), (\ref{eqmx}) and (\ref{eqm}), achieving an
 interesting bridge in particle physics.  Specially interesting is
 the Higgs mass, coming  from a Yukawa-type equation (\ref{eqhx}),
which is smaller by a factor
$\sqrt{\frac{4 \pi}{3 \alpha}}$ than the one derived from a SU(5)
 GUT without gravitation.  Because of the smallness of
$\sqrt{\frac{4 \pi}{3 \alpha}}$ , a contraction period could arise
 in the
 early universe  if $\chi_{o}<0$, otherwise if  $\chi_{o}> 0$, one
 has a normal chaotic kind of inflationary scenario.   On the one
 hand, it  would seem that "chaotic"  initial conditions are more
 appealing, because of their generic
character.  But, on the other hand, it could be also true that
 some unknown, quantum or classical, initial conditions put
 $\chi_{o}$  on a very special value, in consequence of which one has at
first a deflation
 and then automatically the desired inflation as follows from scenario (a).

After inflation, the universe is oscillation dominated, and
without its decay one could explain the missing mass problem of
cosmology given today in the form of cold dark matter.

Our presentation is, however, not free of difficulties: The model, as a whole,
cannot explain immediately the today  observed baryon mass of the universe,
 for which one
is forced to look for a reheating scenario after inflation.
 Perhaps, this must
take place, but the question whether to much gravitational
 radiation is generated to eventually spoil a normal
nucleosynthesis procedure remains open at this energy scale.
 Nevertheless, it is not
necessary to adjust the parameter $\lambda$ to achieve
a density perturbation spectrum required for galaxy formation:
the fact that $\alpha=10^7$ in itself makes
$\delta\rho / \rho \sim \sqrt{\lambda}/\alpha$ reasonable
small, and on the other hand, without attaching the reheat
temperature too low.   We also obtain sufficiently
small gravity perturbations to be in accordance with
the measured anisotropy of the microwave background spectrum.

We think that the "natural" relationship among the fundamental
masses achieved by the theory sheds some light on the understanding
of the present known problems of cosmology.

\vspace{1cm}
 {\bf{Acknowledgments}}

One of the authors (J.L.C.) acknowledges DAAD and CONACyT (reg. 58142)
for the grant received since without it the  work would not have been achieved.

\newpage

\newpage
\begin{center}{\bf  Figure captions  } \end{center}
\bigskip\bigskip

{\bf Figure 1: (a)} The Higgs field come down the hill  from $\chi_{o}<0$.
During
this time the universe contracts, in a "new deflation" scenario with
"fast" rollover evolution of $\chi$. It is possible, however, for very special
initial values that the field in (a) evolves to (b) resulting a
normal exponential
expansion.   {\bf (b)} The Higgs field come down the hill, but now from
$\chi_{o}>0$.  During
that time the universe expands exponentially in a type of a "chaotic"
inflationary scenario.

\bigskip\bigskip

{\bf Figure 2:}   The scale factor $a(t)$  is shown for both inflationary
models (a) and (b) in a logarithmic scale.  The model (a) begins with a
"fast" contraction followed automatically by an inflation  if
$\chi_{o} \approx
-0.15509$.  The upper curve (scenario (b)) shows the behavior of inflation if
$\chi_{o} \approx 130/3$ (chaotic exponential expansion).

\bigskip\bigskip

{\bf Figure 3:}  The Higgs field of scenario (a) as a function of time.
The Higgs field
goes first very fast until it reaches $\chi {}^{>}_{\sim} 130/3
(\approx 280 |\chi_{o}|)$;
at that point $H$ evolves faster than $\chi$, to proceed with an
inflationary phase.

\bigskip\bigskip

{\bf Figure 4:}  The same as in figure (3) but here with initially
$\chi_{o} {}^{>}_{\sim} 130/3 $ . The exponential expansion takes
place directly.

\bigskip\bigskip

{\bf Figure 5:} Again the scale factor evolution as in figure 2, but now till
$t=10^{2} M^{-1}_{H}$.  One notes that the inflation time is approximately
$t=2 ~10^{-37} sec$, later on, the universe is "dark" matter dominated,
perhaps until today, if reheating didn't take away the coherent
Higgs oscillations.  It can be seen the track imprinted by the Higgs coherent
oscillations in the scale factor evolution at that time scale; later on, this
influence will be imperceptible.
\bigskip\bigskip

{\bf Figure 6:} The evolution of the Higgs oscillations is shown in
logarithmic scale during and after inflation.  In scenario (a) the Higgs field
jumps from very small values to $130/3$ to achieve inflation, later it begins
to oscillate.  In scenario (b), the Higgs field disminishes until it begins to
oscillate.

\end{document}